\documentclass{PoS}
\usepackage{amsmath,bm,amssymb,color}

\usepackage{graphicx}
\def\dSigma{{\rm d} \sigma}
\def\dsigma{{\rm d} \hat\sigma}

\title{NNLO QCD corrections for $Z$ boson plus jet production}

\ShortTitle{NNLO QCD corrections for $Z$+jet production}

\author{Aude Gehrmann-De Ridder\\
          Institute for Theoretical Physics, ETH, CH-8093 Z\"urich, Switzerland\\
          Department of Physics, University of Z\"urich, CH-8057 Z\"urich, Switzerland\\
        E-mail: \email{gehra@phys.ethz.ch}}

\author{Thomas Gehrmann\\
        Department of Physics, University of Z\"urich, CH-8057 Z\"urich, Switzerland\\
        E-mail: \email{thomas.gehrmann@uzh.ch}}

\author{Nigel Glover\\
        Institute for Particle Physics Phenomenology, Department of Physics, University of Durham, Durham, DH1 3LE, UK\\
        E-mail: \email{e.w.n.glover@durham.ac.uk}}

\author{Alexander Huss\\
  Institute for Theoretical Physics, ETH, CH-8093 Z\"urich, Switzerland\\
          Department of Physics, University of Z\"urich, CH-8057 Z\"urich, Switzerland\\
        E-mail: \email{ahuss@phys.ethz.ch}}

\author{\speaker{Thomas Morgan}\\
        Institute for Particle Physics Phenomenology, Department of Physics, University of Durham, Durham, DH1 3LE, UK\\
        E-mail: \email{t.a.morgan@durham.ac.uk}}

\abstract{We discuss the next-to-next-to-leading order (NNLO) QCD corrections to $Z$ boson production in association with a jet including all partonic channels at all color levels and including the leptonic decay of the $Z$ boson.  
We focus on the optimization of the numerical evaluation of the double-real contribution and demonstrate that our procedure for spreading the Monte Carlo integration over ${\cal O}(1000)$ cores and recombining the results afterwards lead to stable results with sensible error estimates. We apply representative cuts on the jet and charged lepton transverse momenta and pseudorapidities at LHC energies and present the transverse momentum and
rapidity distributions of the charged leptons.}

\FullConference{12th International Symposium on Radiative Corrections (Radcor 2015) and LoopFest XIV (Radiative Corrections for the LHC and Future Colliders)\\
		15-19 June, 2015\\
		UCLA Department of Physics \& Astronomy Los Angeles, USA}

\begin{document}
\section{Introduction}

The production of $Z$ bosons decaying leptonically is an important process at high-energy hadron colliders such as the Large Hadron Collider (LHC). Pairs of high transverse momentum leptons provide a clean signal at the LHC with a large event rate. Resonant pair production is a key process calibrating parts of the detector and for measuring electroweak and strong interaction parameters. It is therefore vital that the theoretical predictions can match the precision of the high quality experimental data available for this process.

At the LHC, the $Z$ boson is usually produced with additional QCD radiation. $Z$-boson production in association with a hadronic jet is important for Standard Model phenomenology, providing a range of precision measurements to test our understanding of electroweak and strong interaction physics~\cite{ATLASZJ,CMSZJ}. $Z$+jet production measurements can also be used to fit the gluon PDF and the strong coupling $\alpha_s$ in certain energy regimes.

To improve the precision and accuracy of the theoretical prediction for the $Z$+jet process, a variety of theoretical corrections have been considered. The next-to-leading order corrections (NLO) corrections to $Z$+jet production have been known for some time~\cite{ZJNLO}. Furthermore, the NLO EW corrections have also been calculated~\cite{ZJNLOEW}, including multijet merging~\cite{ZJNLOEWM}. Recently we presented the first calculation of the QCD corrections to $Z$+jet at next-to-next-to-leading order (NNLO) precision~\cite{ZJNNLO}. 
In the original work, we presented results for the $qg$, $\bar{q}g$, $q\bar{q}$ and $gg$ at leading color accuracy. We have since updated our calculation to include all partonic channels at all color levels. Our results are obtained using a parton-level event generator that provides the corrections in a fully differential form.

A second calculation of $Z$+jet production at NNLO precision has been presented by Boughezal et al.~\cite{ZJBoughezal} using a different subtraction scheme. In coordination with the authors of~\cite{ZJBoughezal}, we performed an in-depth comparison, by running our code with their settings (cuts, parton distributions, scale choice). This comparison uncovered an error in the numerical code used in~\cite{ZJBoughezal}, which alters their published results. After correction of this error, the code developed in~\cite{ZJBoughezal} agrees with our results. 

\section{Antenna subtraction and regulating IR divergences}

In QCD, the inclusive cross section, after renormalization and mass factorization, takes the factorized form,
\begin{equation}
\dSigma = \sum_{ij} \int_0^1 \frac{d\xi_1}{\xi_1} \frac{d\xi_2}{\xi_2} f_i(\xi_1,\mu^2_F)f_j(\xi_2,\mu^2_F) \dsigma_{ij}(\alpha_s(\mu_R),\mu_R,\mu_F),
\end{equation}
where the probability of finding a parton of type $i$ in the proton, carrying a momentum fraction $\xi$, is described by the PDF $f_i(\xi,\mu^2_F)d\xi$. $\dsigma_{ij}$ denotes the partonic cross section for parton $i$ to scatter off parton $j$, which is summed over all possible initial state partons and convoluted with the PDFs.

For high scattering energies, the partonic cross section can be calculated using a perturbative expansion in $\alpha_s$,
\begin{equation}
\dsigma_{ij} = \dsigma_{ij}^{\mathrm{LO}} + \left ( \frac{\alpha_s(\mu_R)}{2\pi}\right ) \dsigma_{ij}^{\mathrm{NLO}} + \left ( \frac{\alpha_s(\mu_R)}{2\pi}\right )^2 \dsigma_{ij}^{\mathrm{NNLO}} + \mathcal{O}(\alpha_s^3),
\end{equation}
where $\dsigma_{ij}^{\mathrm{NLO}}$ and $\dsigma_{ij}^{\mathrm{NNLO}}$ denote the NLO and NNLO corrections to the process.

In this expansion, all contributions must be infrared (IR) finite. However, it is well known that both the real and virtual matrix elements at a given order contain IR divergences. These singularities conspire to cancel each other to form the finite observable cross section. This cancellation is not immediately obvious given that the real and virtual matrix elements are integrated over phase spaces of different dimensionality. To obtain an IR finite cross section prediction, a subtraction scheme must be used to shift the singularities between the phase spaces in such a way that the IR~pole cancellation is explicit.

\begin{figure}[t]
(a)\includegraphics[width=0.3\textwidth]{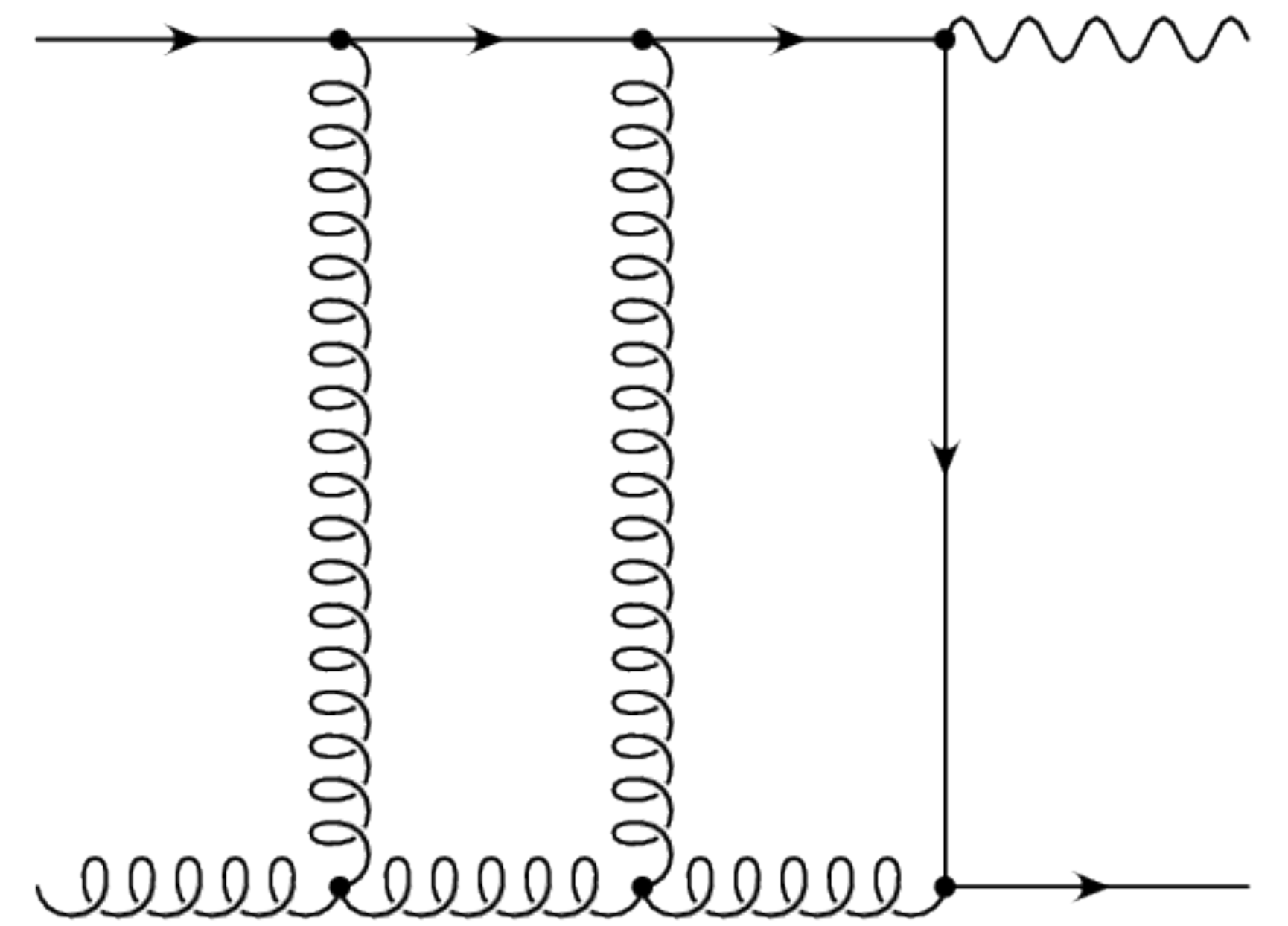}
(b)\includegraphics[width=0.3\textwidth]{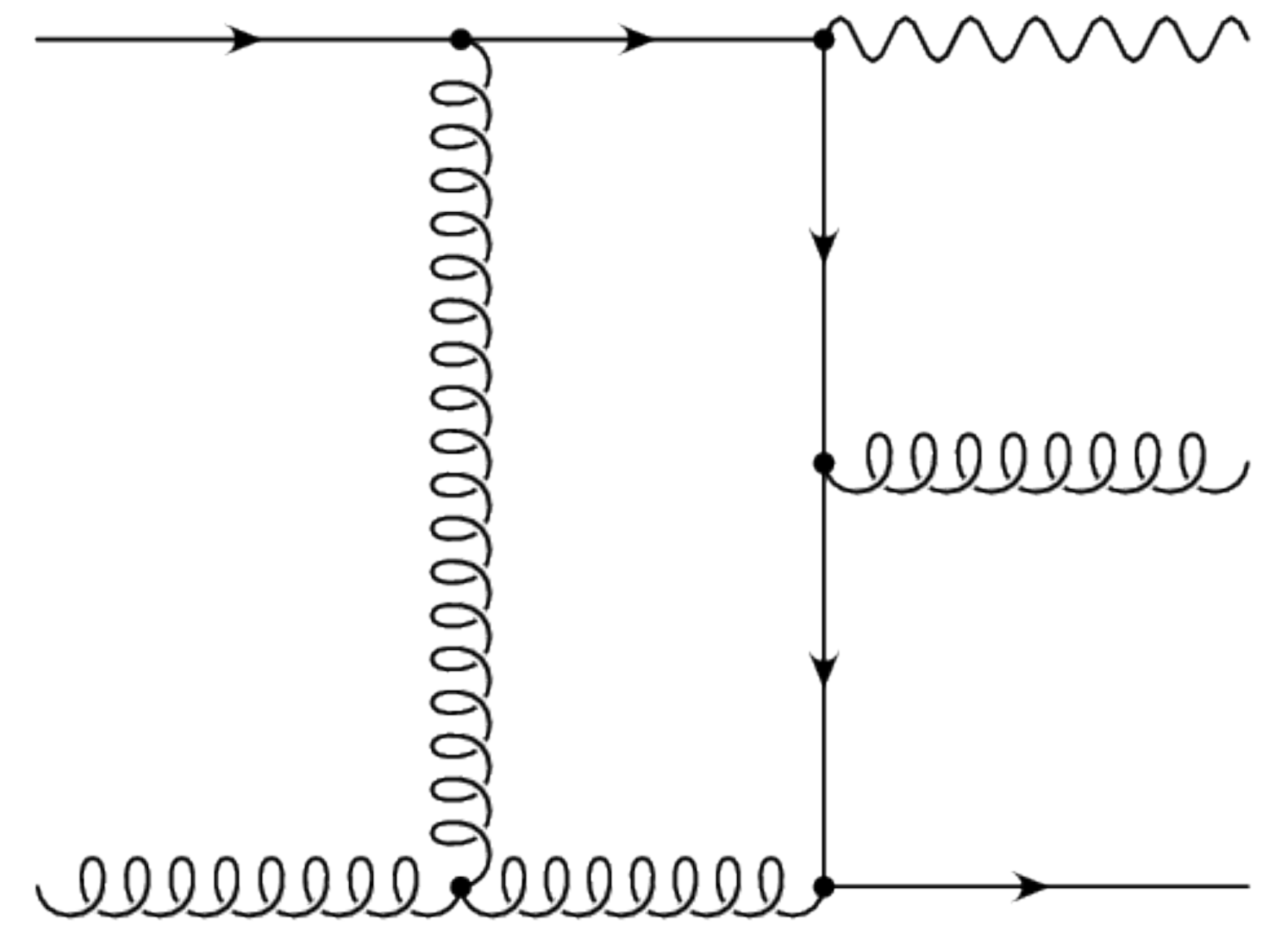}
(c)\includegraphics[width=0.3\textwidth]{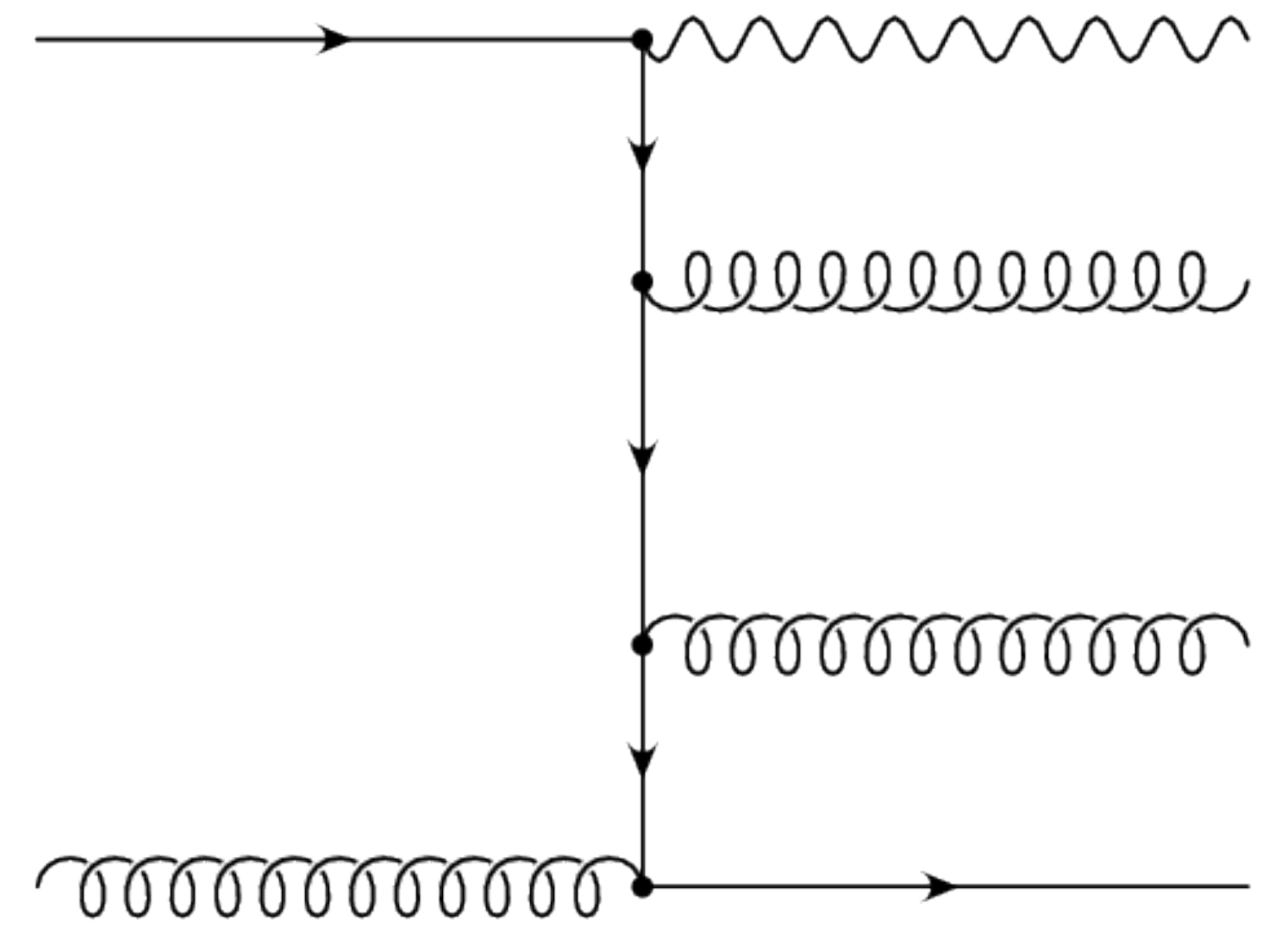}
\caption{Representative Feynman diagrams for  
(a) two-loop $Z$ boson-plus-three-parton amplitudes (b)
one-loop $Z$ boson-plus-four-parton amplitudes and (c) 
tree-level $Z$ boson-plus-five-parton amplitudes.\label{fig:FD}}
\end{figure}

The NNLO corrections to $Z$-boson + jet production receives contributions from three types of parton-level processes: 
(a) $\dsigma_{ij,NNLO}^{VV}$, the two-loop corrections to $Z$-boson-plus-three-parton processes~\cite{Z3p2l}, 
(b) $\dsigma_{ij,NNLO}^{RV}$, the one-loop  corrections to $Z$-boson-plus-four-parton processes~\cite{Z4p1l,ZJJNLO} and 
(c) $\dsigma_{ij,NNLO}^{RR}$,  the tree-level $Z$-boson-plus-five-parton  processes~\cite{Z5p0l,ZJJNLO}. 
Figure~\ref{fig:FD} shows representative Feynman diagrams for each of the partonic multiplicities.

At NNLO, a variety of subtraction schemes including sector decomposition~\cite{secdec}, sector-improved residue subtraction~\cite{stripper}, $q_T$-subtraction~\cite{qtsub}, project to Born subtraction~\cite{Cacciari}, antenna subtraction~\cite{ourant} and $N$-jettiness subtraction~\cite{njettiness} have been successfully applied to the calculation of NNLO corrections for a range of LHC processes. Our results are presented using the antenna subtraction formalism, where the real radiation singularities are described using 3- and 4-parton antenna functions that capture the IR behaviour of the unresolved radiation emitted between a pair of hard radiator partons. Our implementation is based around splitting the integral according to the phase space multiplicity,
\begin{eqnarray}
\dsigma_{ij,NNLO}&=&\int_{{\rm{d}}\Phi_{3}}\left[\dsigma_{ij,NNLO}^{RR}-\dsigma_{ij,NNLO}^S\right]
\nonumber \\
&+& \int_{{\rm{d}}\Phi_{2}}
\left[
\dsigma_{ij,NNLO}^{RV}-\dsigma_{ij,NNLO}^{T}
\right] \nonumber \\
&+&\int_{{\rm{d}}\Phi_{1}}\left[
  \dsigma_{ij,NNLO}^{VV}-\dsigma_{ij,NNLO}^{U}\right],
\label{eq:cross}
\end{eqnarray}
where each of the square brackets is IR finite and well behaved in the infrared singular region. Powerful checks of our formalism are that (a) the $\epsilon$-poles cancel analytically with the pole structure of the one-loop and two-loop matrix elements and (b) the subtraction terms accurately reproduce the singularity structure of the real radiation matrix elements.

\section{Numerical evaluation of the double-real contribution}

Each additional parton in the phase space introduces three additional degrees of freedom to the problem so that the double-real contribution to the cross section requires integration of the highest dimensionality phase space. Including the decay of the $Z$ boson increases the dimensionality further.  The volume of integration is also significantly larger than for an NLO real contribution of equivalent dimensionality (i.e.\ the real contributions to the NLO $Z$+2 jet rate) because it is necessary to integrate over all double unresolved emissions. 
For the numerical evaluation, we use a phase-space generator that is highly optimized for the antenna subtraction formalism and perform the integration using the {\tt VEGAS}  algorithm.  
For the double-real phase space we use ${\cal O}(10^7)$ events to warm up the {\tt VEGAS} grid followed by ${\cal O}(10^8)$ events for the production runs (which produce the total cross section and distributions).  

To achieve this, the ${\cal O}(10^8)$ events can be (a) run on a single core (taking an unreasonably long wallclock time) or (b) split across many cores.   To split across many cores, we can use (i) MPI%
\footnote{{http://www.mcs.anl.gov/research/projects/mpi/}} 
(requires specific code modifications and limited to a single cluster) (ii) OpenMP%
\footnote{{http://openmp.org}} 
(limited by number of cores sharing the resources of a single node) or (iii) as $N$ independent runs across $M$ physically disconnected cores.   

We adopt the latter option, dividing the $N\times M \sim{\cal O}(10^8)$ events in runs of $M\sim{\cal O}(10^5)$ events spread across $N\sim{\cal O}(10^3)$ cores and exploiting the LHC Computing Grid. In this case, we have to be very careful to make a statistically valid treatment of the datasets when combining runs. 

When $M$ is small, each individual run (of $M$ events) does not necessarily provide an accurate estimate of the total cross section since individual events can dominate the cross section for that particular run, leading to an ``outlier'' evaluation of the cross section.   Such outlying events can arise when the phase space point for the full momentum configuration is above a physical cut on an observable, for example a cut on $p_T$, and the subtraction term is evaluated just below the cutoff. This will generate an unusually large weight which must be appropriately taken into consideration. Within a differential distribution, outliers can be formed when the observable evaluated for the matrix element does not lie in the same bin as the subtraction-term observable. This is particularly prominent in rapidity distributions.  

For a single run (of $M$ events) the unweighted average of the function $f(\vec{x})$ is given by
\begin{equation}
\sigma_i(M) = \sum_{j=1}^{M} \frac{f(\vec{x}_j)}{M}
\end{equation}
with corresponding unweighted statistical uncertainty $\delta\sigma_i(M)$.
The weighted average of $N$ such runs is given by,
\begin{equation}
\sigma(N,M) = \frac{\sum_{i=1}^N \omega_i(M)\sigma_i(M)}{\sum_{i=1}^N \omega_i(M)},
\end{equation}
where $\omega_i(M)$ is the weight of a particular run $i$, given by
\begin{equation}
\omega_i = \frac{1}{\delta\sigma_i^2(M)}.
\end{equation}
The importance of the weighted average lies in the effective suppression of outliers since a run containing outliers is typically accompanied with a larger error and, thus, enters the weighted average with a smaller weight $\omega_i$.
In particular, given a fixed number of integrand evaluations ($N\times M$ fixed):
\begin{itemize}
\item For large $M$, small $N$, the $\sigma_i(M)$ have a Gaussian distribution but  the impact of outliers is not taken care adequately, leading to a large uncertainty $\delta\sigma(N,M)$. 
\item For small $M$, large $N$, the $\sigma_i(M)$ do not have a Gaussian distribution and are dominated by outliers but the statistical weighting reduces the effect of outliers leading to a small uncertainty $\delta\sigma(N,M)$.  In this case, we can find a wrong value for  $\sigma(N,M)$ but with a small  $\delta\sigma(N,M)$.
\end{itemize}
To optimize the efficiency of the numerical integration, a compromise between $N$ and $M$ must be found while guaranteeing the correctness of the final answer. 

\begin{figure}
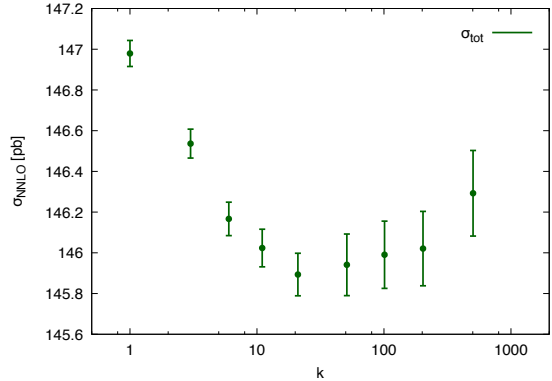

\centering
\includegraphics[width=0.5\textwidth]{{{NNLOFullyz.compXS3chan}}} 
\caption{The total cross section $\sigma(N/k, k\cdot M)$ at NNLO as a function of $k$, the number of runs combined to form a pseudorun. The cross section is then the weighted average of the pseudoruns. The errors on the individual data points are statistical. \label{fig:wgt}}
\end{figure}

We first consider the total cross section, and show $\sigma(N/k, k\cdot M)$ as a function of $k$ in Fig.~\ref{fig:wgt}.
In other words, we form the unweighted average of $k$ runs to form a pseudorun. The cross section is then the weighted average of the pseudoruns. 
As expected, as $k$ increases, the statistical error grows. 
On the other hand, as $k$ decreases, we see a sizeable shift in the central value which amounts to a difference of almost 1\% at $k=1$ with respect to the value of the cross section in the stable region.

\begin{figure}
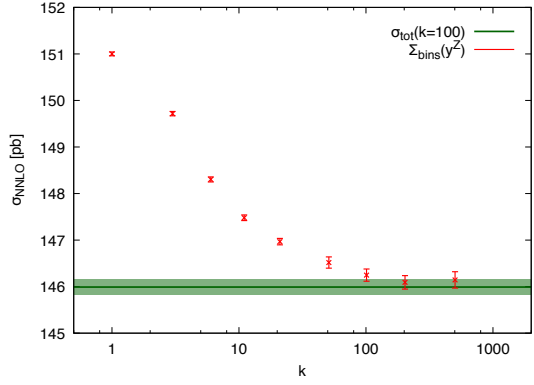

\centering
\includegraphics[width=0.5\textwidth]{{{NNLOFullyz.compXS4chan}}} 
\caption{The integral of the $y^Z$ distribution, $\sigma(N/k, k\cdot M)$ at NNLO as a function of $k$. The errors on the individual data points are statistical. The green band denotes the total cross section evaluated at $k = 100$ with its statistical uncertainty. \label{fig:wgtdist}}
\end{figure}

As a consistency check,  Fig.~\ref{fig:wgtdist} shows the total cross section computed from the integral of the $Z$-boson rapidity distribution as a function of $k$. Similarly to Fig.~\ref{fig:wgt}, there is a sizeable $k$ dependence.  In particular, for $k < 100$ the statistical errors are small, but the central value is off by a large amount.
For $k > 100$, the central value is stable, but the statistical errors increase.

\begin{figure}
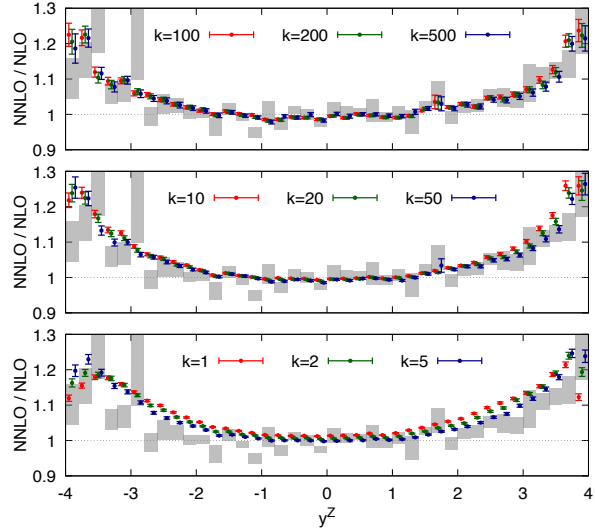

\centering
\includegraphics[width=0.55\textwidth]{{{yz.compKchan}}}
\caption{The ratio of the NNLO against the NLO prediction for individual bins of the rapidity distribution of the $Z$ boson for different values of $k$. The grey bands are the unweighted result. The errors on the individual data points are statistical. \label{fig:yzcomp}}
\end{figure}

We can also consider the effect of combining runs on the $Z$-boson rapidity distribution itself. Fig.~\ref{fig:yzcomp} shows that as $k$ increases, the bin-by-bin statistical errors also increase (as expected). For $k < 5$, we see that the central values for different values of $k$ disagree by a sizeable amount, resulting in an incorrect value for the total cross section evaluated from the integral of this distribution, as shown in Fig.~\ref{fig:wgtdist}.

\begin{figure}
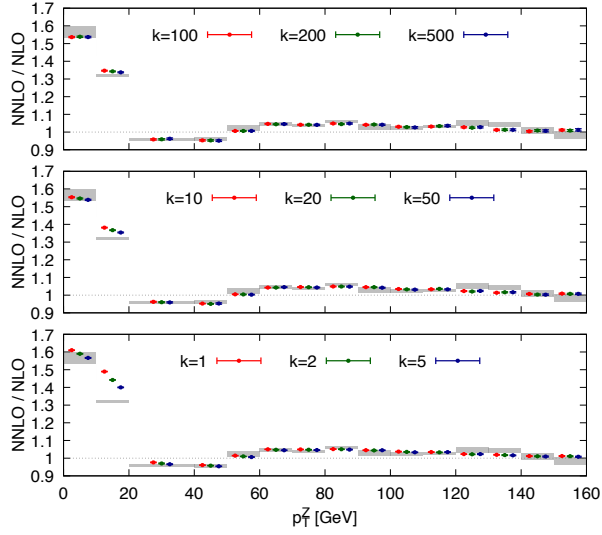

\centering
\includegraphics[width=0.55\textwidth]{{{ptzmrg.compKchan}}}
\caption{The ratio of the NNLO against the NLO prediction for individual bins of the $p_T$ distribution of the $Z$ boson for different values of $k$. The grey bands are the unweighted result. The errors on the individual data points are statistical. \label{fig:ptzcomp}}
\end{figure}

Similar conclusions can be drawn from transverse momentum distribution of the $Z$ boson shown in Fig.~\ref{fig:ptzcomp}. We observe that above the $p_T^{\rm jet}$ cut, the $p_T^{Z}$ distribution is stable as a function of $k$, even at low values. This demonstrates that there are fewer outliers in this distribution compared to the rapidity distribution of the $Z$ boson, as expected, and our errors are more reliable at small $k$.

Below the $p_T^{\rm jet}$ cut we see a sizeable shift in the central value for small $k$. This can be understood because this region probes configurations where the extra partonic radiation present at NLO and NNLO can compensate the transverse momentum of the leading jet, generating a $Z$-boson transverse momentum below the $p_T^{\rm jet}$ cut---the well known Sudakov shoulder phenomenon~\cite{sudakov}. This unusual configuration is very sensitive to the jet definition and is prone to generating outliers because we are close to the physical cuts.

We conclude from these results that a safe value for $k$ is ${\cal O}(100)$, corresponding to $\mathcal{O}(10^7)$ events per pseudorun. For these $k$-values the total cross section (computed directly and from integrating the distributions) and the distributions themselves are insensitive to the precise choice of $k$. We know that our final results are not being shifted by outliers and that our statistical uncertainties are under control.  We therefore use $k=100$ for all the differential distributions presented below and also including those in Ref.~\cite{ZJNNLO}. 

\section{Physical cross sections}

Finally, we extend the results presented in Ref.~\cite{ZJNNLO} to include a breakdown according to the initial state, and exploit the fact that our calculation includes the full decay and spin correlations of the vector boson decay into leptons to give NNLO predictions for the physically observable lepton transverse momentum and pseudorapidity distributions. 
Our studies are based on the same centre-of-mass energy of $\sqrt{s} = 8$ TeV, same input parameters and same experimental cuts as found in Ref.~\cite{ZJNNLO}. We use the PDF set NNPDF2.3~\cite{nnpdf} with a corresponding value of $\alpha_s(M_Z)$ = 0.118 at NNLO. This same PDF set is used for the LO, NLO and NNLO predictions with the same value of $\alpha_s(M_Z)$ throughout. The renormalization and factorization scales are chosen to be $\mu = \mu_R = \mu_F = M_Z$ and the theoretical uncertainty estimated by varying this scale choice by a factor in the range [1/2, 2].

To simulate representative experimental cuts, we require that the leptons have pseudorapidity, $|\eta^{\ell}| < 5$ and that the dilepton invariant mass is close to the $Z$ boson mass, $80$~GeV $< m_{\ell\ell} < 100$~GeV. Jets are reconstructed  using the anti-$k_T$ algorithm~\cite{antiKT} with $R=0.5$ and are required to have $p_T^{\rm jet} >30$~GeV and $|y^{\rm jet}| < 3$.

\begin{table}
\centering
  \begin{tabular}{c|c|c|c}
    \hline
    & \multicolumn{3}{|c}{cross section (pb)} \\
    \hline
  channel & LO & NLO & NNLO \\
  \hline
  \hline
  $qg$            & $53.6^{+4.5}_{-4.3}$   & $80.2^{+3.5}_{-3.2}$  & $84.0^{+2.0}_{-3.9}$ \\
  $q\bar{q}$      & $27.1^{+1.5}_{-1.5}$   & $33.1^{+0.2}_{-0.6}$  & $32.0^{+1.5}_{-2.0}$ \\
  $\bar{q}g$      & $22.9^{+1.7}_{-1.7}$   & $33.1^{+0.0}_{-0.4}$  & $34.9^{+2.2}_{-3.3}$ \\
  $gg$            &   N/A                  & $-4.0^{+3.3}_{-1.9}$  & $-7.2^{+5.1}_{-3.8}$ \\
  $qq$            &   N/A                  & $1.8^{+2.7}_{-1.8}$   &  $2.0^{+3.2}_{-2.6}$ \\
  $\bar{q}\bar{q}$&   N/A                  & $0.1^{+0.4}_{-0.3}$   &  $0.1^{+0.4}_{-0.4}$ \\
  \hline
    total         & $103.6^{+7.7}_{-7.5}$  & $144.4^{+9.0}_{-7.2}$ & $145.8^{+0.0}_{-1.2}$ \\
    \hline
  \end{tabular}
  \caption{Channel breakdown of the total cross section for LO, NLO and NNLO for the scale choice $\mu~=~\mu_F=\mu_R=M_z$. The theoretical uncertainty on each channel is estimated from the envelope of the [1/2, 1, 2]$M_Z$ scale choices.\label{table:channel}}
\end{table}

We see from Table~\ref{table:channel} that for each channel there are relatively small shifts in their contribution to the total cross section going from NLO to NNLO.  However, the most significant shift is in the $gg$-initiated channel which receives a correction of nearly 80\% compared to the NLO prediction. This significant shift is not surprising as this channel first appears at NLO. The $gg$ contribution lies mainly around $y^Z \sim 0$ and will be highly relevant for fitting the gluon PDFs using the $Z$+jet LHC data. 

\begin{figure}
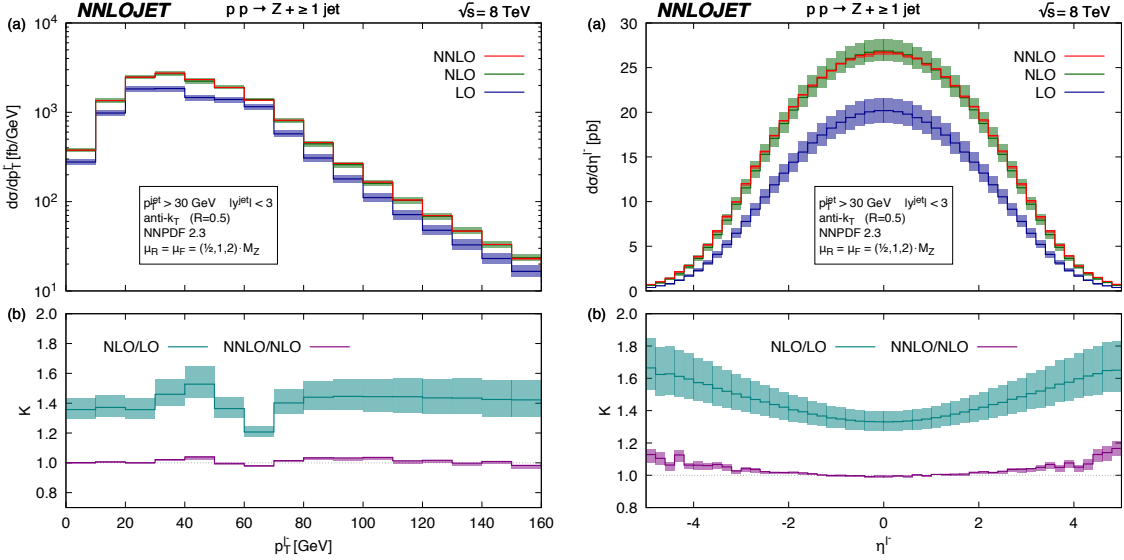

\includegraphics[width=0.5\textwidth]{{{ptlm}}}
\includegraphics[width=0.5\textwidth]{{{ylm}}}
\caption{Transverse momentum distribution (left) and pseudorapidity distribution (right) for the positively charged lepton for $Z(\to \ell^+\ell^-)$+jet production in $pp$ collisions with $\sqrt{s} = 8$~TeV at LO (blue), NLO (green), NNLO (red).
The lower panels show the relative ratios of the perturbative orders, NLO/LO (turquoise) and NNLO/NLO (mauve).\label{fig:lepton}}
\end{figure}

Figure~\ref{fig:lepton} (left) shows that the NNLO corrections to the transverse momentum distribution for the leptons are uniform and small across the entire range in $p_T$. Figure~\ref{fig:lepton} (right) shows that the NNLO corrections to the pseudorapidity distribution for the leptons have a clear shape relative to the NLO prediction. We see that in the central regions the NNLO and NLO predictions agree very well, however at large pseudorapidities the NNLO corrections are sizeable with up to 10\% corrections over the NLO prediction.

\section{Conclusions}

In conclusion, we have discussed the complete NNLO QCD calculation of $Z$-boson production in association with a jet in hadronic collisions including all partonic subprocesses and the leptonic decay of the $Z$ boson.  
We demonstrated that our procedure for combining different sets of parallelized numerical integration runs is fully under control in our code, both for total cross sections and for distributions. We extended the results presented in Ref.~\cite{ZJNNLO} to include a breakdown according to the initial state, and exploited the fact that our calculation includes the full decay and spin correlations of the vector boson decay into leptons to give NNLO predictions for the physically observable lepton transverse momentum and pseudorapidity distributions.  

\section*{Acknowledgements}
This research was supported in part by the Swiss National Science Foundation (SNF) under contracts 200020-162487 and CRSII2-160814, in part by
the UK Science and Technology Facilities Council as well as by the Research Executive Agency (REA) of the European Union under the Grant Agreement PITN-GA-2012-316704  (``HiggsTools''), and the ERC Advanced Grant MC@NNLO (340983).

\end{document}